\documentclass[12pt]{iopart}
\usepackage{graphicx}
\begin{document}
\title{Functionalizing graphene by embedded boron clusters}
\author{Alexander Quandt$^1$\footnote{Permanent address:
Institut f{\"u}r Physik der Universit{\"a}t Greifswald,
Felix--Hausdorff-Str.\ 6, 17489 Greifswald, Germany.}, Cem
\"{O}zdo\u{g}an$^2$, Jens Kunstmann$^3$, Holger Fehske$^4$}
\address{$^1$CNR--IFN Trento, CSMFO group, Via Alla Cascata 56/c, 38050
Povo--Trento, Italy.}
\address{$^2$ Department of Computer Engineering, \c{C}ankaya
University, Balgat, 06530 Ankara, Turkey.}
\address{$^3$ Max--Planck--Institut f{\"u}r Festk{\"o}rperforschung,
Heisenbergstrasse 1, 70569 Stuttgart, Germany.}
\address{$^4$ Institut f{\"u}r Physik der Universit{\"a}t Greifswald,
Felix--Hausdorff-Str.\ 6, 17489 Greifswald, Germany.}
\ead{quandt@physik.uni-greifswald.de}

\date{\today}
\begin{abstract}
We present a model system that might serve as a blueprint for the
controlled layout of graphene based nanodevices. The systems
consists of chains of {\rm B}$_7$ clusters implanted in a graphene
matrix, where the boron clusters are not directly connected. We
show that the graphene matrix easily accepts these alternating
boron chains, and that the implanted boron components may
dramatically modify the electronic properties of graphene based
nanomaterials. This suggests a functionalization of graphene
nanomaterials, where the semiconducting properties might be
supplemented by parts of the graphene matrix itself, but the basic
wiring will be provided by alternating chains of implanted boron
clusters that connect these areas.
\end{abstract}
\submitto{\NT}
\maketitle

%
%


Single layers of graphite called graphene are the precursors of
carbon nanotubes \cite{dresselhaus96}, which are one of the key
materials for nanotechnology. However, since the discovery of
stable multilayers and single layers of graphene
\cite{novoselov04}, the latter quickly shifted into the focus of
nanotechnology as well. For carbon nanotubes a simple tight
binding scheme \cite{saito98} predicts that they should either be
semiconducting or metallic, depending on their chirality
\cite{dresselhaus96}. The same theory predicts that stripes cut
from bulk graphene may either be semiconducting or metallic,
depending on the nature of their boundaries
\cite{nakada96}\footnote{These standard results were verified
using a MAPLE 9.5 worksheet developed by one of us (A.\ Q.).}.

But a broad semiconducting sheet of graphene could be a suitable
basis on which to built an ultimate two-dimensional nanoelectronic
technology. This would require a controlled wiring of such devices
on the nanoscale, which has little to do with the clumsy
state--of--the--art wiring of graphene devices made for the
purpose of basic scientific investigations. In fact the tiny
wiring necessary for technologically relevant nanoelectronic
applications of graphene must primarily be based on a chemistry
that is largely compatible with carbon chemistry, and the purpose
of this paper is to suggest such a system.

In the field of nanotubes, it has been shown that boron nanotubes
are largely compatible with carbon nanotubes \cite{quandt05}. In
particular, interfaces between boron and carbon nanotubes were
shown to be stable \cite{kunstmann04}. The basis for this
compatibility must be sought in the electron deficient nature of
boron, which leads to a complex, but somewhat adjustable
multi--centered bonding. Another favorable property of electron
deficient elements like boron is their ability to force other
elements into unusually high coordinations \cite{pauling60}. This
property should allow for a certain range of structural
adjustments to be made whenever boron fragments are in touch with
a carbon environment, as seen in the case of nanotubular
interfaces \cite{kunstmann04} or mixed boron--carbon clusters
\cite{exner00}.

Finally it turns out that small boron clusters are quasi--planar,
and largely made from pyramidal {\rm B}$_7$-units
\cite{boustani97}. The same building blocks are the basis of novel
boron nanomaterials, including boron nanotubes or boron sheets
\cite{quandt05, kunstmann06}, the latter being the boron
equivalent of graphene. Interesting enough, all bulk boron
nanomaterials made from these {\rm B}$_7$-units are metallic. Now
from a pure geometrical point of view, a quasi--planar boron
cluster made from {\rm B}$_7$-units would fit quite perfectly into
the honeycomb lattice, as a carbon honeycomb may also be seen as
the basis of a hexagonal pyramidal {\rm C}$_7$-unit, where the
carbon at the top of the carbon pyramid has been removed.

But pure geometric arguments are not sufficient to establish boron
as a suitable wiring component for graphene based
nanotechnologies. We must also require chemical compatibility
between the carbon matrix and implanted boron fragments. To this
end, we have developed a suitable model system, and studied its
basic structural and electronic properties using ab initio
methods.

\begin{figure*}[hb]

\begin{tabular}{ll}
\includegraphics[scale=1,width=8cm]{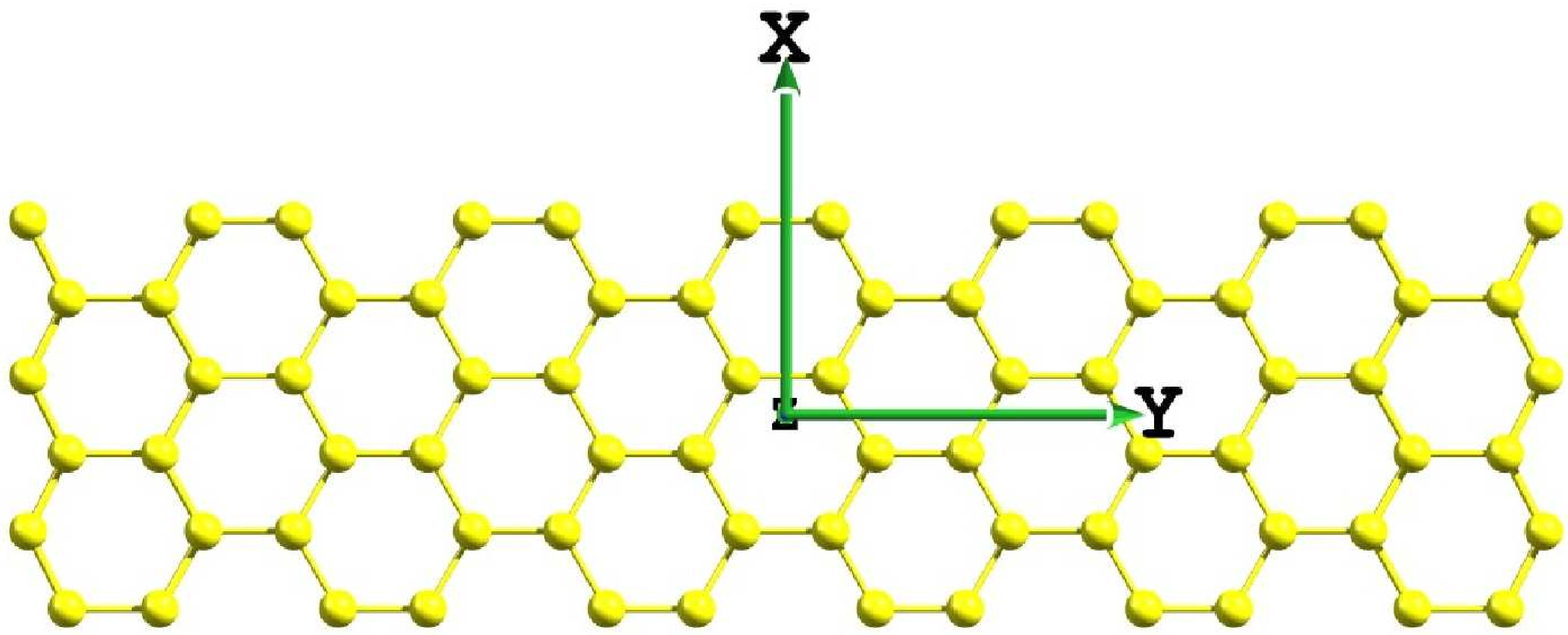} & \includegraphics[scale=0.23]{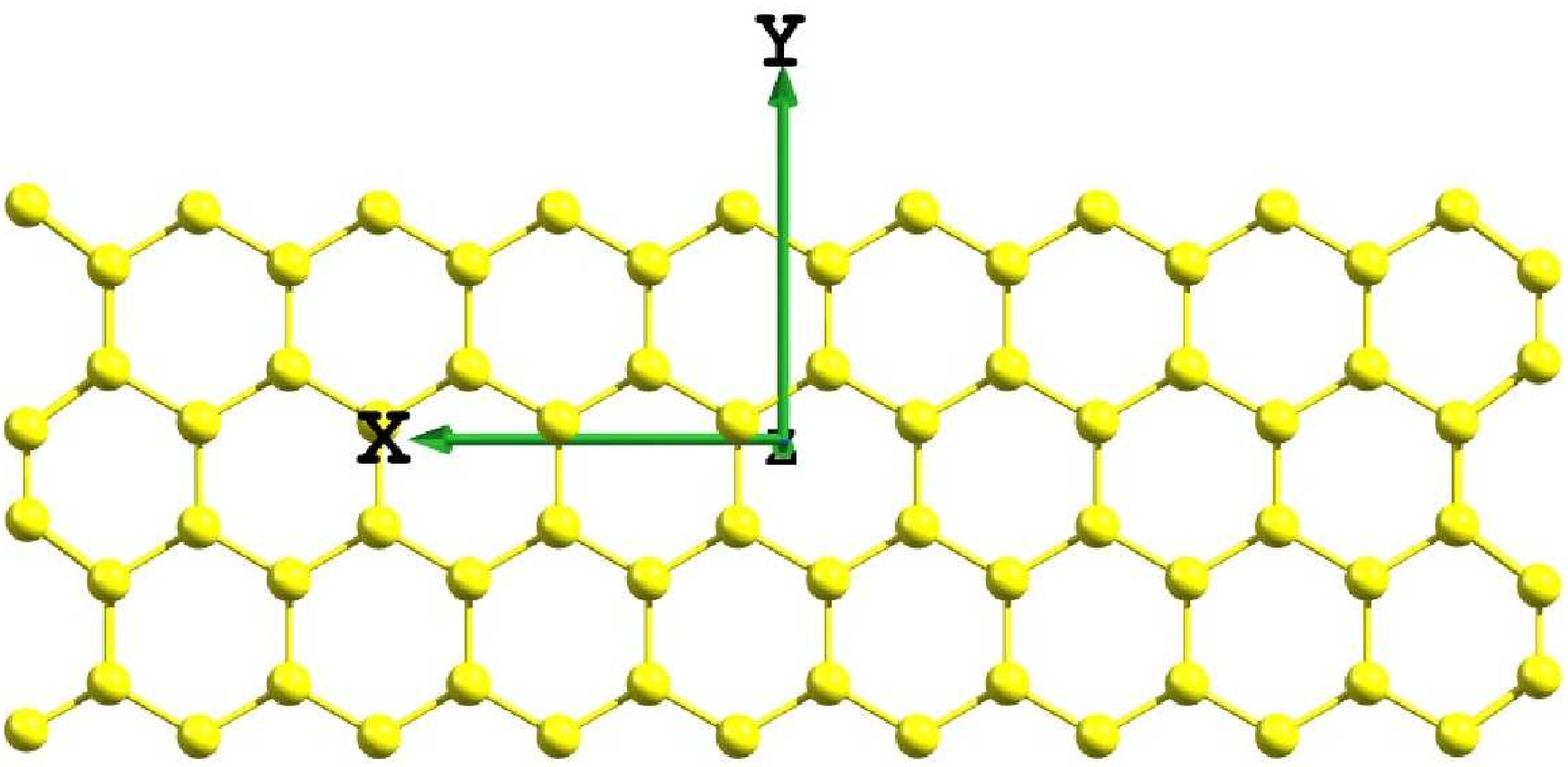}\\
{\bf{(a)}} & \bf{(b)}\\
\includegraphics[scale=0.24,angle=-90]{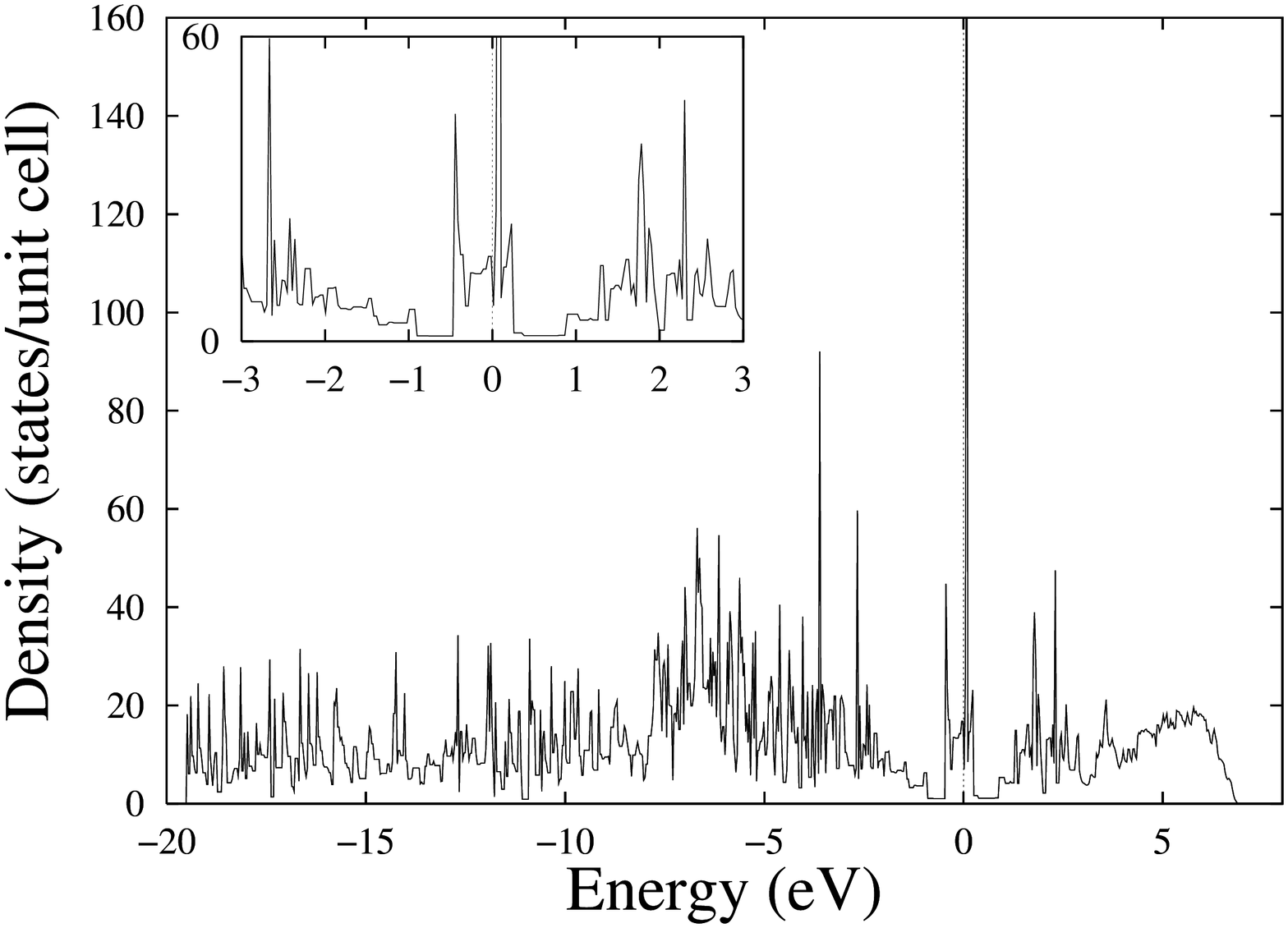} & \includegraphics[scale=0.24,angle=-90]{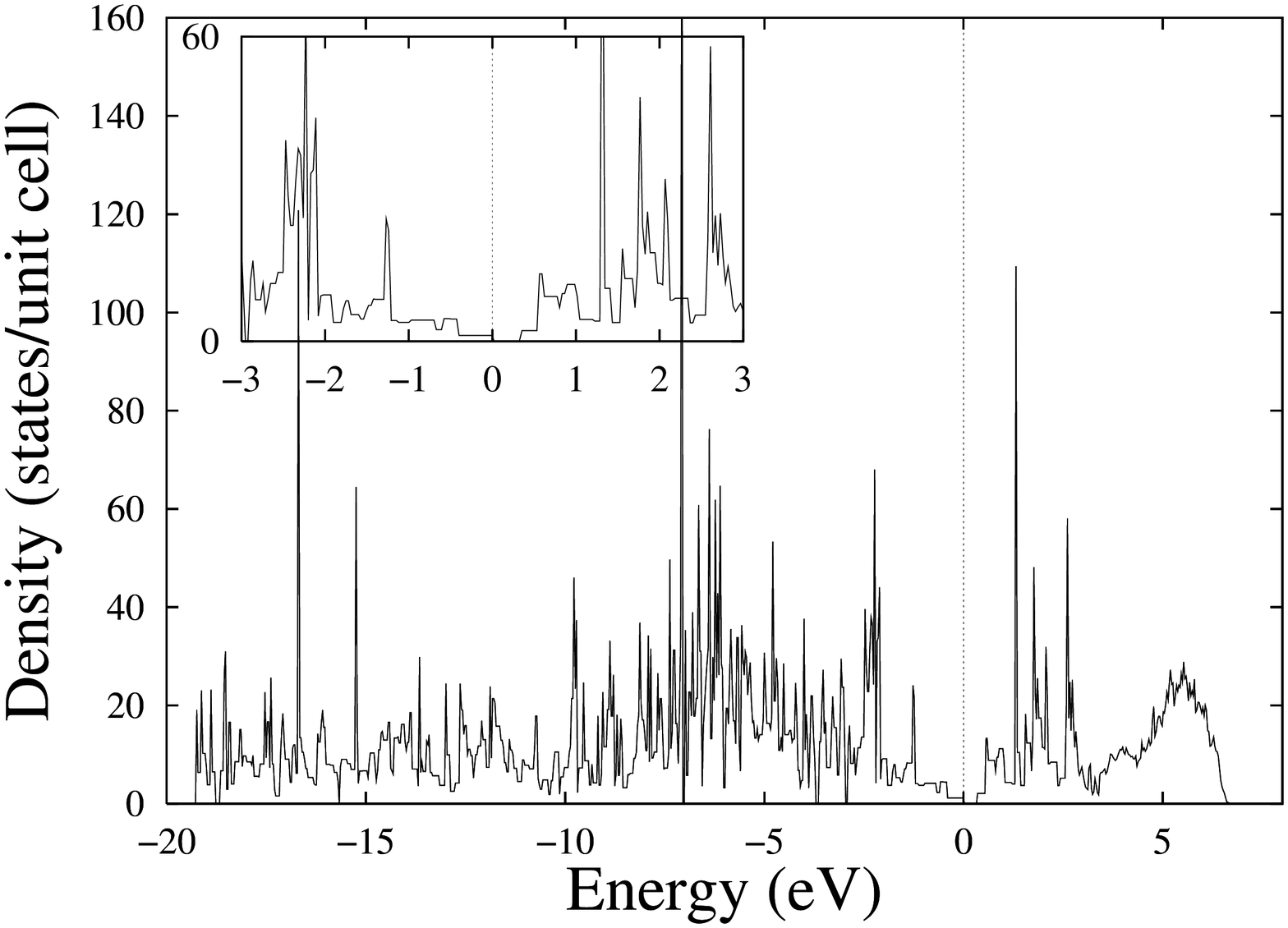}\\
\bf{(c)} & \bf{(d)}\\
\includegraphics[scale=0.22,angle=-90]{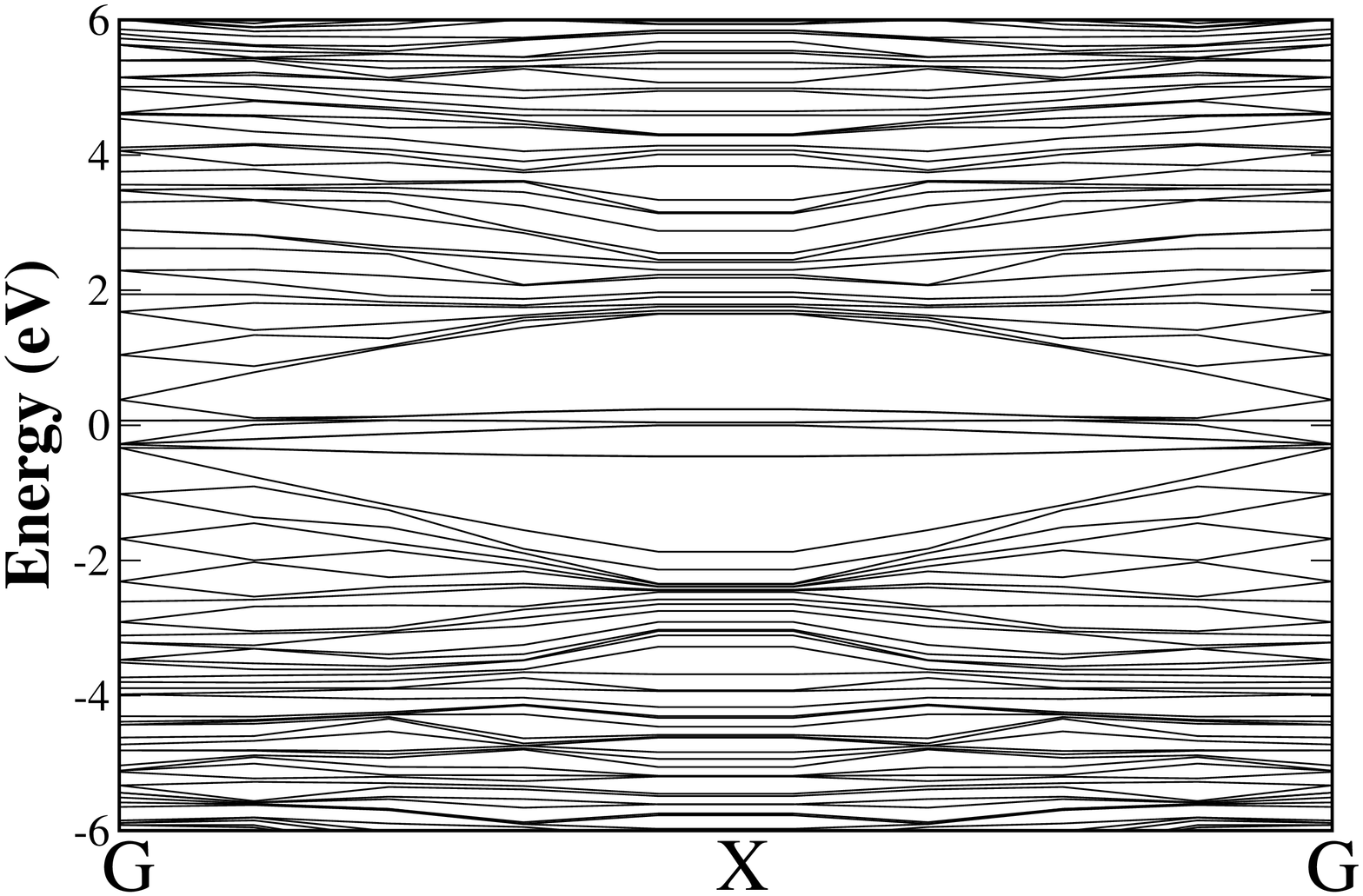} &  \includegraphics[scale=0.22,angle=-90]{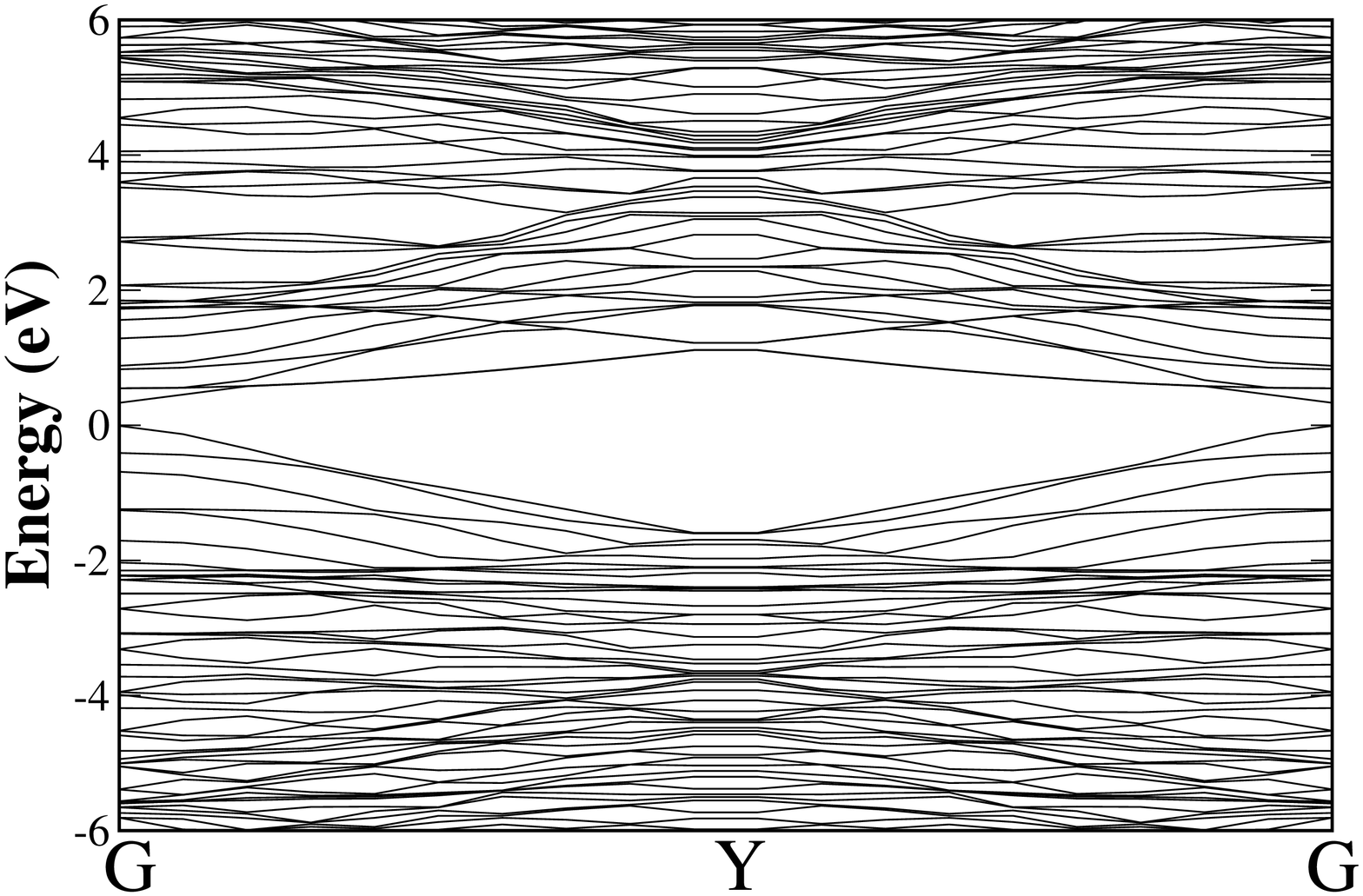}\\
\bf{(e)} & \bf{(f) }
\end{tabular}
\caption{\label{fig1} Graphene nanoribbons. Figure shows relaxed
structure of (a) zigzag nanoribbon and (b) armchair nanoribbon,
where the horizontal direction spans the width of the ribbon, and
the vertical direction corresponds to the periodic direction of
the systems. Density of states for (c) zigzag nanoribbon that
shows metallic behavior, and for (d) armchair nanoribbon that
shows semiconducting behavior, as predicted by tight binding
theory \cite{saito98, nakada96}. The insets show details of the
density of states around the Fermi level located at 0 eV, and the
Fermi level has been emphasized by a vertical bar. Electronic
bands are drawn in the $k_x$-direction for (e) zigzag nanoribbon,
and in the $k_y$-direction for (f) armchair nanoribbon. $G$
denotes the Gamma point of the Brillouin zone.}
\end{figure*}

\begin{figure*}[hb]

\begin{tabular}{lll}
\multicolumn{3}{l}{
\includegraphics[width=10cm]{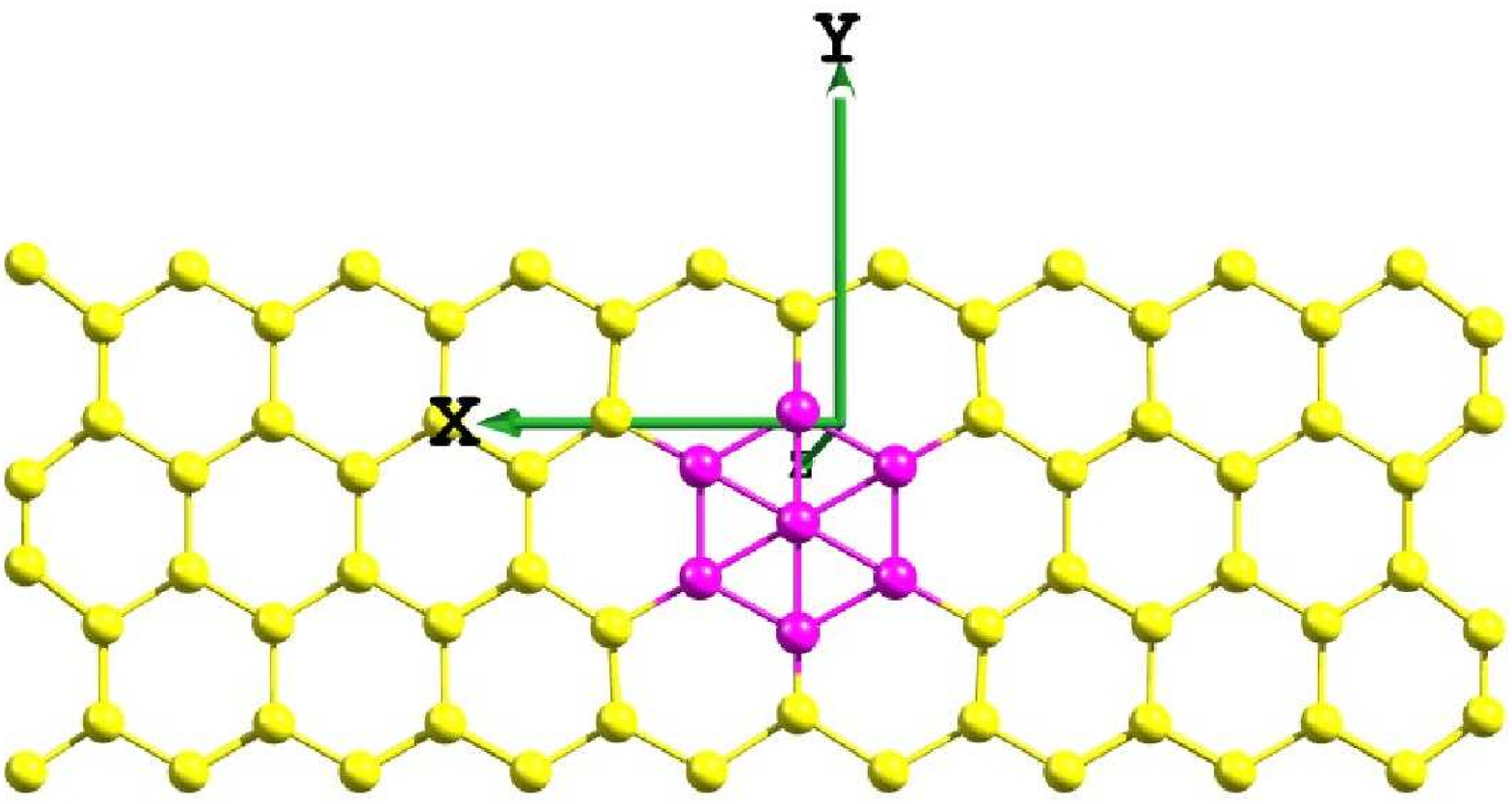}} \\
\multicolumn{3}{l}{ \bf{(a)} } \\
\includegraphics[scale=0.23,angle=-90]{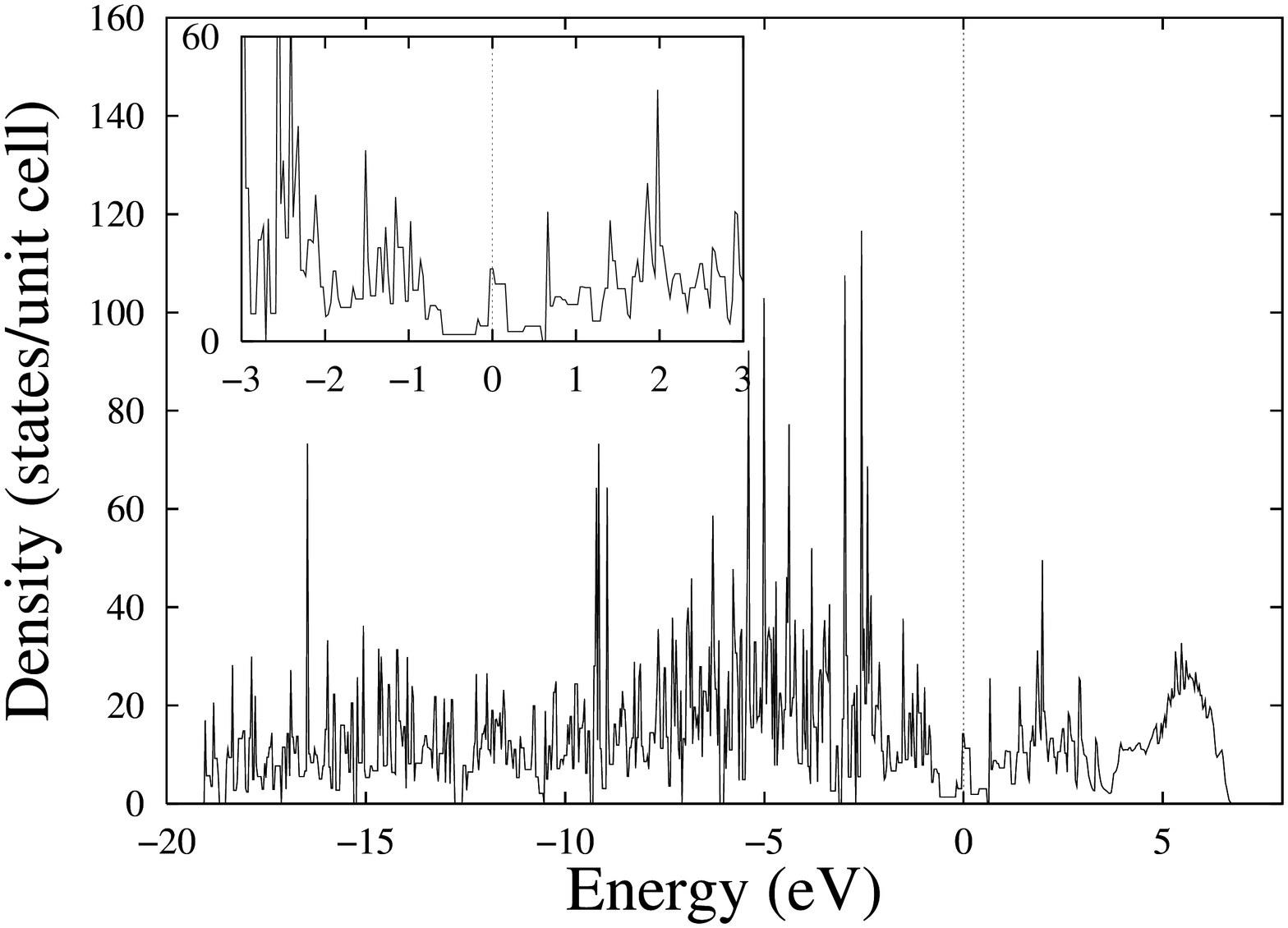} &
\multicolumn{2}{l}{
\includegraphics[scale=0.21,angle=-90]{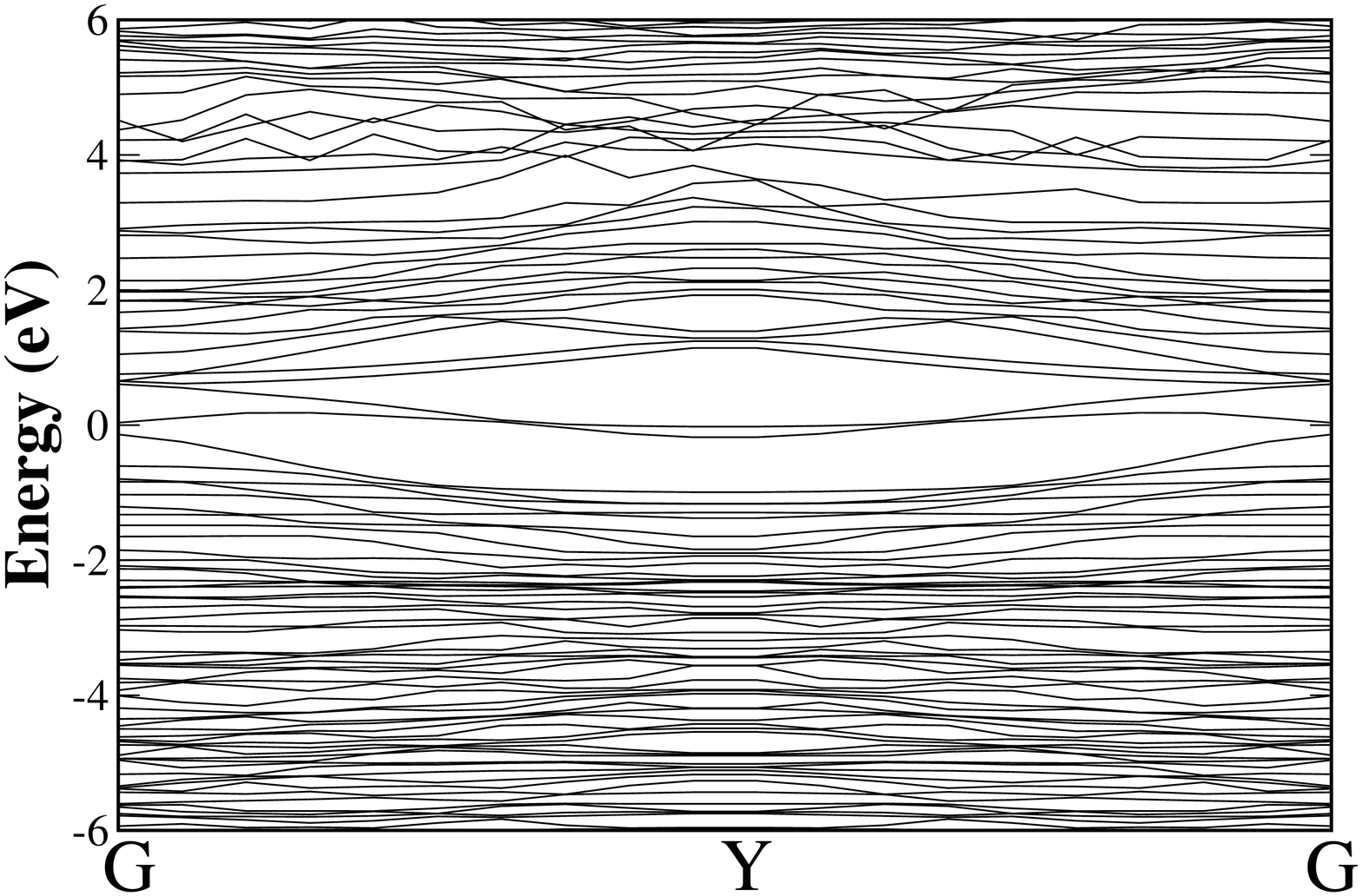}} \\
\bf{(b)} & \multicolumn{2}{l}{\bf{(c)} } \\
\includegraphics[scale=0.23,angle=-90]{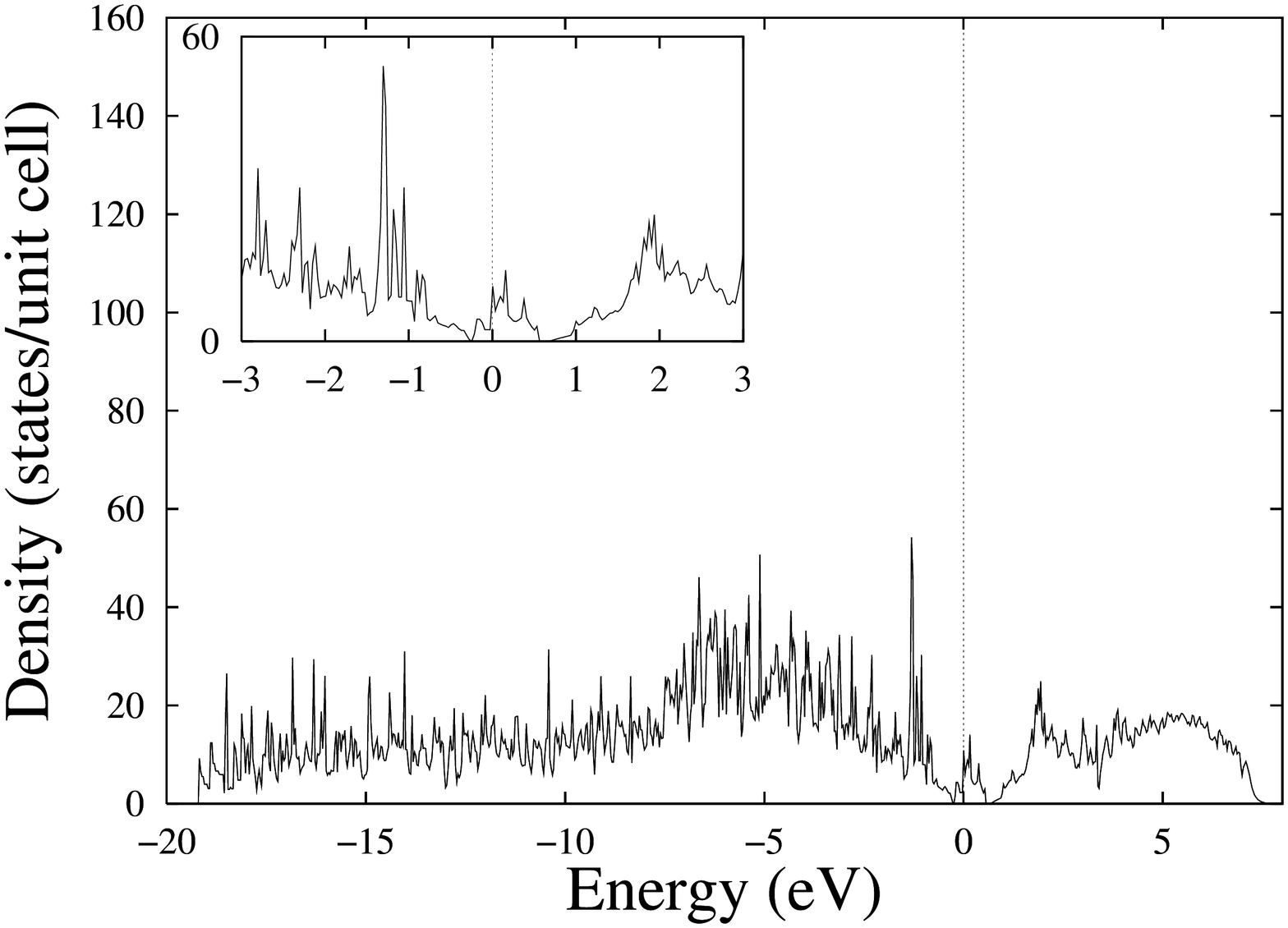} &
\includegraphics[scale=0.21,angle=-90]{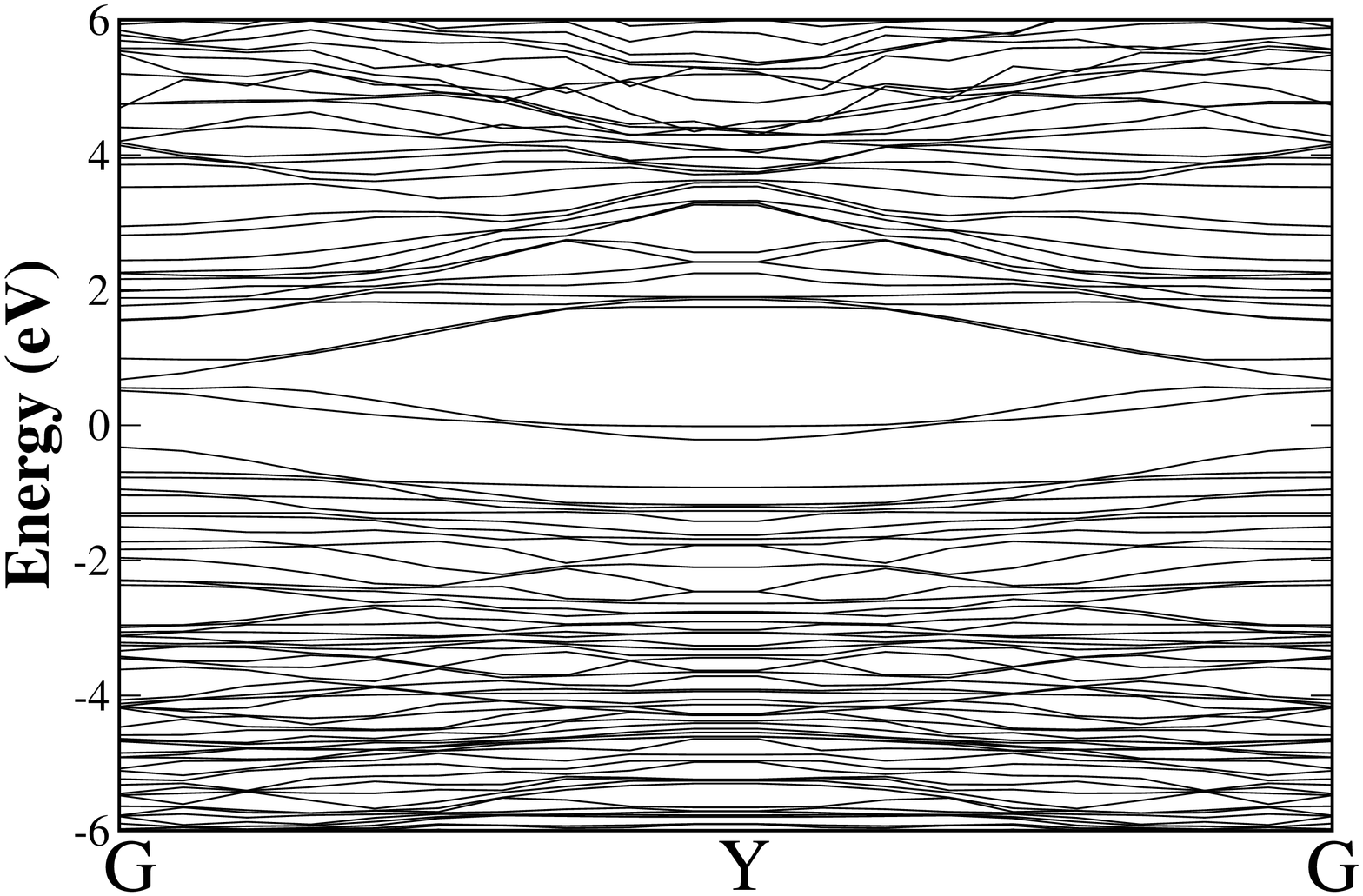} &
\includegraphics[scale=0.21,angle=-90]{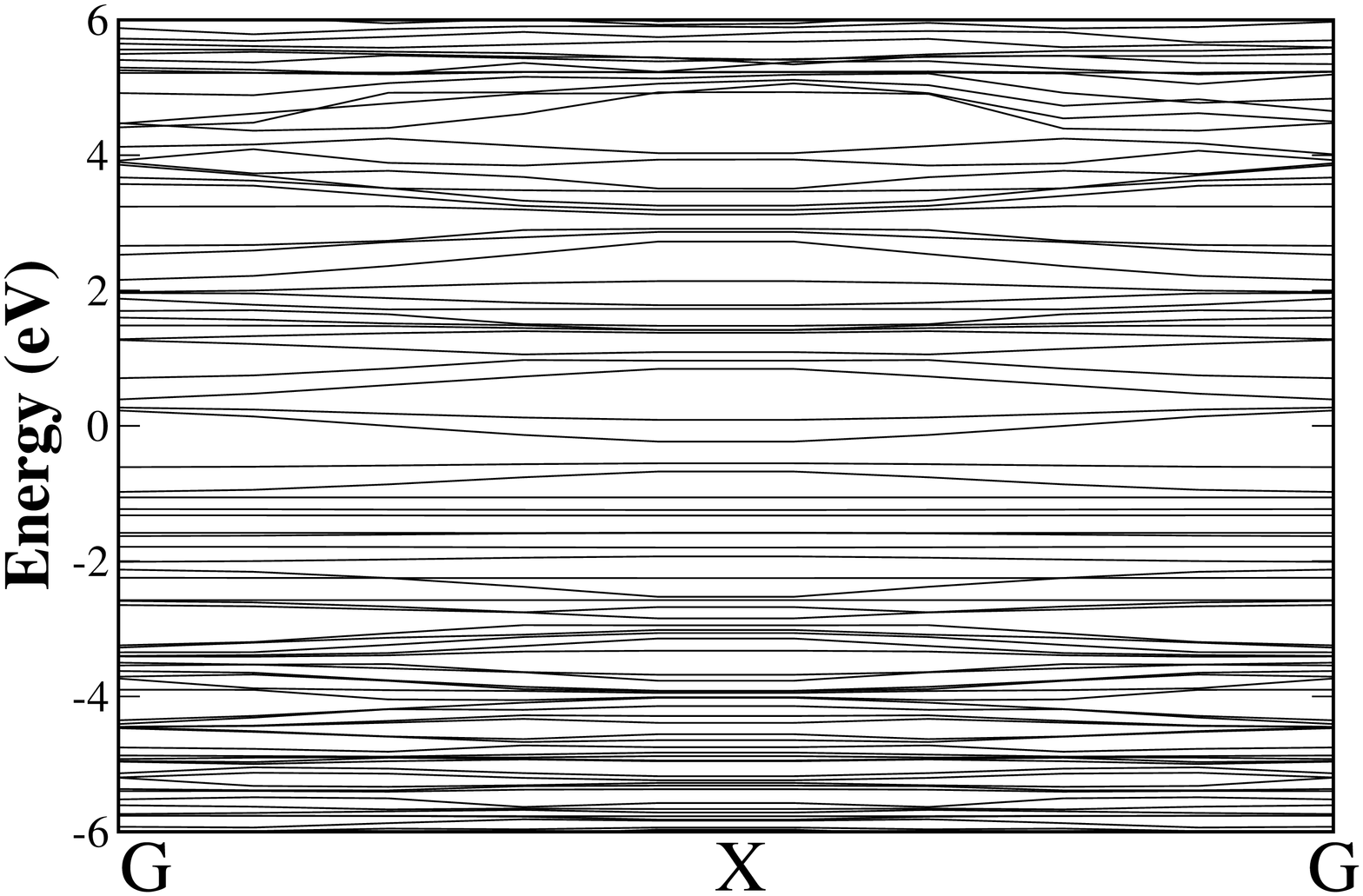} \\
\bf{(d)} & \bf{(e)} & \bf{(f) }
\end{tabular}
\caption{\label{fig2} Boron doped graphene nanostructures. (a)
Relaxed structure of boron doped armchair graphene, where the
vertical $y$--direction corresponds to the periodic direction of
the system. This structure is characterized by an alternating
chain of \rm{B}$_7$ clusters (shown in dark) along the
$y$-direction, separated by a carbon honeycomb. The relaxed
structure of boron doped graphene looks rather similar, but it is
periodic in both the horizontal ($x$) and the vertical ($y$)
direction. (b) Density of states for doped armchair graphene that
shows metallic behavior. (c) Electronic bands in the
$k_y$-direction that show the corresponding band overlap at the
Fermi level. (d) Density of states for boron doped graphene,
similar to (b). (e)-(f) Electronic bands in the $k_y$-- and
$k_x$--direction for boron doped graphene. }
\end{figure*}

These ab initio calculations and optimizations were carried out
using the VASP package (version 4.6.28) \cite{kresse96-1,
kresse96-2}, which is a density functional theory \cite{kohn65}
based ab initio code using plane wave basis sets and a supercell
approach to model solid materials, surfaces, or clusters
\cite{teter89}. In our case, where all the systems should
effectively be two-dimensional (i.e. monolayers), we were choosing
supercells that allowed for a huge distance at right angles to
these layers (i.e. the $z$--direction in Figs.\ \ref{fig1} and
\ref{fig2}). For the nanoribbons considered later on, we would
also choose another huge distance in the direction of the width of
those nanoribbons.

During our simulations, the electronic correlations were treated
within the local-density approximation (LDA) using the
Perdew-Zunger \cite{perdew81} form of the Ceperley-Alder
exchange-correlation functional \cite{ceperley80}, and the ionic
cores of the system were represented by ultrasoft pseudopotentials
\cite{vanderbilt90} as supplied by Kresse and Hafner
\cite{kresse94}. The k--space integrations were carried out using
the method of Methfessel and Paxton \cite{methfessel89} in first
order, where we employed a smearing width of 0.1 eV. The optimal
sizes of the k--point meshes for different systems were
individually converged, such that changes in the cohesive energy
were reduced to less than 3 meV/atom.

The structures were fully relaxed including lattice parameters and
ions. In order to wash out the energy landscape in search for
global minima, and in order to prevent our quasi two-dimensional
systems from collapsing into multi--layered configurations during
structure optimization, we would run VASP for a number of ionic
steps with reduced precision (using only half of the converged
k--point meshes), followed by a reset of the huge lattice
parameters of the unit cell. This procedure was repeated several
times.

Following these pre--relaxations, we would restrict the structure
optimizations to comprise the ionic degrees of freedom, only, but
this time we used the optimal size of the k--point
meshes\footnote{Optimized k--point meshes for graphene (10x10x3),
zigzag ribbon 1 (6x3x3) and 2 (3x3x3), armchair ribbon 1 (2x5x2)
and 2 (2x3x3), doped armchair ribbon (2x3x3) and doped graphene
(2x3x3), using the notation of table \ref{table1}.}. The optimized
systems would finally undergo a series of static calculations,
where we applied the tetrahedron method \cite{bloechl94} for
k--point sampling. At this stage, we would check that the stress
on the system and the interatomic forces were sufficiently small,
in order to obtain accurate cohesive energies listed in table
\ref{table1}, and accurate band structures shown in Figs.\
\ref{fig1} and \ref{fig2}. The density of states of the systems
shown in Figs.\ \ref{fig1} and \ref{fig2} were obtained in a
static run using four times the optimal size of the k--point
meshes, in order to guarantee the proper resolution of small
details.

In figure \ref{fig1}, we see the results of such simulations for
graphene nanoribbons with zigzag and armchair borders. In the
vertical direction of figure \ref{fig1}(a) and (b), the systems
repeat themselves periodically, whereas in the horizontal
direction, the ribbons just span the width shown in these
pictures, and a neighboring nanoribbon to the right or to the left
of it will be too far away to interact with this strip. Note that
the smallest unit cells for these systems would comprise only one
third of the atoms shown in the zigzag case, and one half of the
atoms shown in the armchair case. However we extend the unit cells
in order to be sure that our general procedures would remain valid
in the size ranges that we would finally choose for the doped
model to be discussed below. The near identical cohesive energies
for all settings, and the comparison of these energies with the
cohesive energy of graphene listed in table \ref{table1}, were a
strong indication that our basic procedure for carrying out ab
initio simulations on such systems was reliable.

Tight--binding theory \cite{saito98, nakada96} predicts all zigzag
nanoribbons to be metallic, and this result is confirmed by the
results shown in figure \ref{fig1}(c) and (e). In the case of an
armchair nanoribbon the same theory predicts all of these
structures to be semiconducting, unless the number of dimers from
one border of the nanoribbon to the other is $3n-1$, with $n$
being a natural number. We see however that the number of dimers
in figure \ref{fig1}(b) is 18, and therefore this nanoribbon
should be semiconducting, which is confirmed by the results
displayed in figure \ref{fig1}(d) and (f).

With the semiconducting armchair system shown in figure
\ref{fig1}, we found an ideal starting point for setting up a
model system that contains (alternating) chains of boron clusters
running in the periodic direction of a semiconducting nanoribbon.
But the question was, which kind of system should be examined in
the first place. One might certainly think of starting with single
boron atoms laid out in the periodic direction. Such studies have
already been carried out quite recently \cite{martins07}, and the
authors observed that single boron atoms have a strong tendency to
diffuse to the borders of the nanoribbons, similar to boron
migration in open-ended boron--doped carbon nanotubes
\cite{hernandez00}. Thus, although this type of doping might be
very important for spin transport through graphene nanoribbons,
which seems to be mediated by edge states, it was of little help
for our goal to find a model that would describe the bulk
functionalization of a broad semiconducting graphene nanoribbon.

Therefore we decided to insert something as large as boron
clusters made from pyramidal {\rm B}$_7$--units, which might not
diffuse that easily to the open ends of a graphene nanoribbon.
Note that for microelectronic devices close to the lithographic
limit and beyond, there is a similar trend towards the
implantation of boron clusters instead of boron atoms
\cite{iit06}, because with shrinking system sizes, single boron
atoms become too mobile, and the MOSFET devices will degrade too
quickly. There is no reason to assume that this situation would be
any better in the case of graphene based devices.

To avoid these problems, the necessary implantation machines for
boron clusters are already available (see \cite{iit06}). Therefore
a first step towards the realization of our model system could be
to shine an intense beam of boron clusters on graphite, pull off
the graphene layers in a standard fashion \cite{novoselov04}, and
start to analyze the resulting patterns. Depending on the
intensity of that cluster source, there might be copies of roughly
the same pattern in neighboring graphene layers. The first results
might be nothing more than similar copies of large speckle
patterns, but the technique could certainly be refined, step by
step.

With such a procedure in mind, we decided to make our model system
slightly more realistic by assuming that the layout of small boron
clusters might not be too precise. Therefore the resulting boron
chains could easily be interrupted by carbon honeycombs, thus
forming alternating boron chains running across a suitable bulky
graphene nanostructure. This leads us to the model system shown in
figure \ref{fig2}(a), where a {\rm B}$_7$--cluster has been
implanted in the interior of the armchair nanoribbon of figure
\ref{fig1}, and the implanted boron clusters within neighboring
unit cells are separated by a carbon honeycomb. Note that if we
reduce the huge horizontal distance in the $x$--direction between
neighboring nanoribbons, such that the left border of the
nanoribbon shown in figure \ref{fig2}(a) would come into bonding
distance with its right border (periodic boundary conditions), we
obtain a structure model for boron doped bulk graphene, where
parallel lines of alternating cluster chains run in the vertical
$y$--direction of figure \ref{fig2}(a), but their mutual distances
in the horizontal $x$--direction are rather large.

We would like to emphasize that the configuration shown in figure
\ref{fig2}(a) is not the starting configuration, but already the
final relaxed configuration for a boron doped nanoribbon. The
final structure of boron doped graphene is almost identical to
this picture, and therefore we renounced on depicting this
structure in figure \ref{fig2}. The fact that the starting
configurations and the optimized configurations are so similar may
certainly be taken as a strong indication, that the basic
chemistry of both components must be compatible. Furthermore, the
cohesive energies shown in table \ref{table1} for these system are
in a range that one would expect from a stable implantation of
small boron cluster into bulk graphene nanomaterials.

The most interesting properties of these compound system are of
course their basic electronic properties. In figure \ref{fig2}(b)
and (d) we see that for the boron doped armchair nanoribbon and
for boron doped graphene, the corresponding density of states
predicts a metallic behavior, whereas in figure \ref{fig1}(d) for
the non--doped armchair nanoribbon, we would find an electronic
gap at the same place. The reason for the absence of a gap in the
case of the boron doped systems are two overlapping bands shown in
figure \ref{fig2}(c) and (e). The remaining features of the band
structure and density of states remain similar to the ones for
non-doped armchair graphene shown in figure \ref{fig1}(d) and (f).
This may be taken as an indication that the appearance of Bloch
states at the Fermi level with crystal momentum $\vec{k}$ in a
direction parallel to the alternating boron chains is a rather
robust and dominant effect.

Note that in the case of the doped armchair nanoribbons, the bands
for Bloch states with $\vec{k}$--vectors in the horizontal
$x$--direction (spanning the width of the nanoribbons) are flat,
due to a large lattice parameter chosen to isolate them from
neighboring nanoribbons. But in the case of boron doped graphene,
where those nanoribbons have been shortcircuited, we have a
conduction band pulled down below the Fermi level for Bloch states
with $\vec{k}$--vectors in the horizontal $x$--direction (at right
angles to the direction of the alternating boron chains), which
might be coupled to Bloch states with $\vec{k}$--vectors in the
vertical $y$--direction parallel to the alternating boron chains.
But these features could as well be a reminder of the metallic
behavior of the underlying graphene matrix.

At this point, a naive picture emerges of a nanotechnology with
conducting "dotted" lines of boron clusters running through a
broad piece of graphene, thus connecting parts of these sheets
that act as a basic semiconducting substrate for nanoelectronic
devices. Nevertheless, for various reasons this picture might be
misleading, and therefore a simple study like the present one
certainly needs to be supplemented by further and more detailed
research in the near future.

First of all, with our methods we could not determine the nature
of the states around the Fermi levels shown in figure \ref{fig2}.
They could be extended states, or localized states provided by the
embedded boron clusters. The former would lead to metallic
conductivity, the latter to hopping conductivity. We could even
have a mixed situation of hopping and metallic conductivity along
the alternating cluster chains.

Second, it has recently been shown that spin polarization effects
absent from our LDA calculations may lead to a band splitting for
metallic (zigzag) nanoribbons, thus opening a gap around the Fermi
level \cite{son07}. We therefore repeated some of our calculations
using spin polarized DFT. It turned out that in the case of (boron
doped) armchair nanoribbons, there was no qualitative difference
to the results shown in figure \ref{fig2}, and that is why we
restricted our discussions to results obtained with non polarized
LDA.

Finally there are other fundamental aspects that could not be
treated in the framework of such a simple study, like the
determination of the distance between boron clusters in the
periodic direction, where the metallic behavior might finally
disappear. Other interesting topics could be the implantation of
even larger boron clusters, or the modelling of inhomogeneous or
disordered cluster chains. With little modifications of the
present model, and given sufficient computational resources, all
of these questions might be answered pretty soon.

%
%

\section*{Acknowledgements}
We would like to thank the staff at the HLR Stuttgart for their
assistance during our extensive use of the HLRS supercomputing
facilities, and we are particularly indebted to Dr.\ F.\
G{\"a}hler from ITAP Stuttgart for introducing us into the
handling of the local VASP installations at the HLRS. C.\ {\"O.}
also acknowledges financial support by HPC Europe.

%
%
\newpage
\section*{References}
\bibliographystyle{iopart-num}
\bibliography{litver}
\newpage

%
%
%

\section*{Tables}

\begin{table}[hb]
\caption{\label{table1} Number of atoms per unit cell and cohesive
energies for various doped and non-doped graphene based
nanomaterials. Note that for the zigzag and armchair nanoribbons,
the system with a minimal unit cell labelled by 1 and the system
with a doubled or tripled unit cell labelled by 2 describe the
same physical system, which explains why the cohesive energies are
practically identical.}
\begin{indented}
\item[ ]
\begin{tabular}{lcl}\br
name &number of atoms& {cohesive energy}\\
     &per unit cell & in eV/atom\\
\mr
graphene &2  & -10.133\\

zigzag ribbon 1 &24   &  -9.830\\

zigzag ribbon 2 &72   &      -9.832\\

armchair ribbon 1 &36   &  -9.868\\

armchair ribbon 2 &72   &  -9.864\\

doped armchair ribbon &73 & -9.505\\

doped graphene &73   & -9.758\\
\br

\end{tabular}
\end{indented}
\end{table}
\newpage
\pagebreak
%
%

\section*{Figures}

%
%
%

\end{document}